\documentclass[prl,%superscriptaddress,
showpacs, twocolumn%, preprint
]{revtex4}
\usepackage{amsmath,amssymb,dcolumn,subfigure}
\usepackage{epsfig}

\begin{document}
\title{Fractional hopping-type motion in columnar mesophase
of semi-flexible rod-like particles}

\author{Saber Naderi,$^{1,5}$ Emilie Pouget,$^{2,3}$ Pierre Ballesta,$^{3}$  Paul van der Schoot,$^{1,4,\ddag}$ M. Paul Lettinga,$^{3}$ and Eric Grelet$^{2,}$}
\altaffiliation{grelet@crpp-bordeaux.cnrs.fr\\
$^\ddag$~p.vanderschoot@phys.tue.nl}
\affiliation{$^1$ Faculteit Technische Natuurkunde, Technische
Universiteit Eindhoven, Postbus 513, 5600 MB Eindhoven, Netherlands\\
$^2$ Universit\'{e} de Bordeaux et CNRS, Centre de Recherche Paul-Pascal,
 115 Avenue Schweitzer, F-33600 Pessac, France\\
$^3$ ICS-3, Institut Weiche Materie, Forschungszentrum J\"ulich, D-52425
J\"ulich, Germany\\
$^4$ Instituut voor Theoretische Fysica, Universiteit Utrecht, Leuvenlaan 4, 3584 CE Utrecht, Netherlands\\
$^5$ Dutch Polymer Institute, P.O. Box 902, 5600 AX Eindhoven, Netherlands}

\date{\today}

\begin{abstract}
We report on single-particle dynamics of strongly interacting filamentous \textit{fd} virus particles in the liquid-crystalline columnar state in aqueous solution. From fluorescence microscopy we find that rare, discrete events take place, in which individual particles engage in sudden, jump-like motion along the main rod axis. The jump length distribution is bimodal and centered at half and full particle lengths. Our Brownian dynamics simulations of hard semi-flexible particles mimic our experiments, and indicate that full-length jumps must be due to collective dynamics in which particles move in string-like fashion in and between neighboring columns, while half jumps arise as a result of particles moving into defects. We find that the finite domain structure of the columnar phase strongly influences the observed dynamics.
\end{abstract}

 \maketitle

The interplay between ordering, fluctuations and mobility of particles is a key ingredient in our current understanding of soft matter. The Lindemann criterium for crystal melting purports that a crystal can only be stable provided that fluctuations of the particles away from the equilibrium positions are less than about $1/8$ of the lattice distance \cite{Lindemann1910}, which indeed is observed in real space in crystalline dispersions of spherical colloids \cite{Alsayed2005}. In \textit{glassy} dispersions of colloidal spheres there is no long-range order but particle caging hinders free diffusion: breaking and reformation of cages dominates the dynamics \cite{Kegel2000}.
The situation is more complex for elongated colloidal particles such as DNA, actin filaments and filamentous viruses, because of the additional orientational degree of freedom \cite{Maarel08,Solomon10}. This gives rise to a sequence of mesophases with increasing packing fraction, from the (chiral) nematic where there is only orientational ordering to smectic and (hexatic) columnar liquid-crystalline phases characterized by one- and two-dimensional positional order \cite{Dogic06,GreletPRL08}.

While self-diffusion in  nematic \cite{Lowen99,Bruggen98} and smectic \cite{Bier08,Lettinga07,Pouget11,Cinacchi09,Kuijk12} phases has been investigated theoretically and experimentally, little is known about dynamics in the dense columnar phase, where particles are stacked into liquid-like columns that are organized in a hexagonal array.
Perfect hexagonal ordering is frustrated for the chiral {\it fd} virus by the helical twist of the particles, spawning dislocations. This breaks long-range hexagonal translational order, giving rise to a {\it hexatic} columnar phase \cite{GreletPRL08}. In this paper we report on the intriguing  motion of particles in this highly viscous hexatic phase, using particle tracking of fluorescently labeled {\it fd} viruses. We observe rare and discrete jump events along the director, predominantly of half a particle length, while a significant fraction of events involves full-length jumps. A small fraction of the particles reorient during the jump, hinting at the presence of structural defects, such as dislocations or grain boundaries.

Hopping-type diffusion processes of one rod length have been observed before in the \textit{smectic} of \textit{fd} virus \cite{Lettinga07,Pouget11}. It stems from the underlying lamellar structure of the smectic that has a layer spacing close to one particle length. Note that this length scale is \textit{not} intrinsically present in the columnar phase. Indeed, earlier simulations on a columnar phase of hard sphero-cylinders did not reveal any jumps along the director, only perpendicular to it \cite{Belli10}. This may be due to the very small aspect ratios of the bidisperse rods and the rigidity of the particles \cite{Belli10}. Hence, we performed Brownian dynamics simulations in a single and a multi-column model to investigate different jump scenarios. We find that half jumps may arise following to the creation of a void when a particle leaves a column for a defect, while full jumps seem to be due to highly correlated, circular-type motion of particles when a number of them change columns. %See Figure \ref{Fig_trace_sim}(c).
Both processes are fast and take place in otherwise barely moving collection of particles. In this sense, the dynamics is heterogeneous, i.e., glass-like \cite{Kegel2000,Kang13}.

Before going into the details of the simulations, we present our experimental results. \textit{Fd} viruses (contour length $L=880~nm$, diameter $\sigma=6.6~nm$, and persistence length $P=2.8~\mu m$ \cite{Barry09}) were prepared in aqueous solution having an ionic strength of 20~mM and a pH of 8.2 at a concentration of 130~mg/mL to obtain the columnar mesophase \cite{GreletPRL08}. One virus out of 10$^5$  has been labeled with dyes, allowing for single particle tracking by fluorescence microscopy (See Supplementary Material \cite{SM}) for which the acquisition time was $100~ms$ and the frame rate of one image per $30~s$ to reduce photobleaching.

\begin{figure}
{\includegraphics[width=0.5\textwidth]{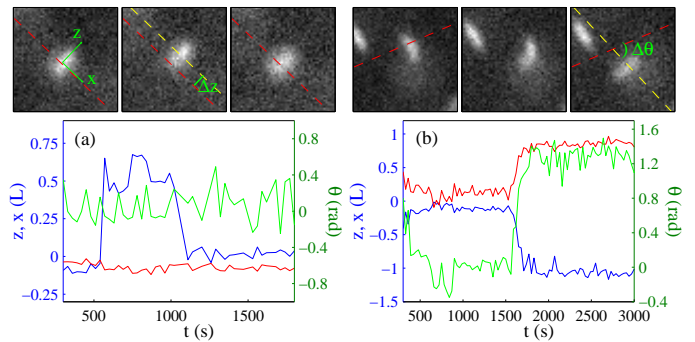}}
\caption{\label{Fig_trace} Hopping-type motion of filamentous {\it fd} virus particles in the hexatic columnar mesophase. In most cases, the particle only jumps (a) but in some others, they also reorient and eventually bend (b). Fluorescence images show the raw data where the red dashed line is drawn perpendicular to the long axis of the labeled particle and indicates its position at initial time. The corresponding traces present the dimensionless displacements along ($z$, blue) and perpendicular ($x$, red) to the long axis (director). Displacements are normalized by the particle length, $L$. The green lines indicate the relative orientation change of the particles in radians (right vertical axes). The experimental uncertainties in determining the position and orientation of the particles are the origin of the noise in the traces.}
\label{Fig_exp}
\end{figure}

\begin{figure}
\includegraphics[width=0.45\textwidth]{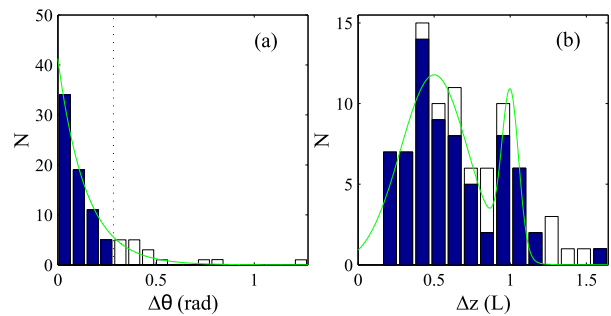}
\caption{\label{Fig_histogram} Experimental distribution of jumps in the columnar mesophase of \textit{fd} virus in aqueous solution.
(a) Rotation angle in radians after a jump event. In white the distribution of jump angles larger than twice the decay angle obtained by an exponential fit (green line). (b) Histogram of the jump length normalized by the rod length, $L$. The green line is a sum of two gaussian functions centered at $\Delta z=L/2$ and $L$.}
\end{figure}

The main experimental result of our investigations is that \textit{rare} events occur in the columnar phase of \textit{fd} virus, where individual particles suddenly and rapidly change their center-of-mass position.
Typical traces are shown in Fig. \ref{Fig_trace}, with the corresponding micrographs. Remarkably, the displacement of the particle shown in Fig. \ref{Fig_trace}(a) amounts to about half a rod length. This fractional displacement is the distinctive feature of our experiments, and represents the majority of the observed events in spite of some jumps of a full particle length (Fig. \ref{Fig_histogram}(b)).

It is important to stress that only about a quarter of the particles engage in a jump event in the three hours time window of the experiment. Assuming that in principle all particles jump, we obtain an estimated jump frequency of about one jump per twelve hours. This is more than two orders of magnitude lower than the hopping-type motion observed in the smectic phase of similar virus suspensions \cite{Lettinga07,Pouget11}. In addition, a significant number of jump events concern particles that simultaneously undergo a large reorientation involving some bending in a process reminiscent of reptation. A typical example is shown in Fig. \ref{Fig_trace}(b), where a particle turns by about $\pi/2$ radians. The distribution of reorientation angles is given in Fig. \ref{Fig_histogram}(a) for which the white bars indicate the particles that undergo a reorientation of more then two times the decay angle. Clearly, large reorientational motion is connected with large displacements, as shown in Fig. \ref{Fig_histogram}(b), and must therefore involve another mechanism than that involved for half jumps.

As the columnar mesophase of the {\it fd} virus suspensions is highly defected due to both its intrinsic short-ranged hexatic features, and the presence of grain boundaries between domains associated with the `powder geometry' of our samples, we conjecture that the reptating or reorienting particles are located near a grain boundary or other type of topological defect such as a screw dislocation \cite{KamienSelinger01}.  This result is consistent with the short-ranged hexagonal ordering measured by X-ray scattering indicating a domain size in the lateral direction of a few rod \textit{diameters} \cite{GreletPRL08}. Note that we ascertained by differential interference contrast microscopy that there are no smectic domains remaining in the columnar phase.

We also studied single-particle dynamics in the columnar phase of {\it fdY21M} virus. Compared to wild-type {\it fd}, this mutant is both less chiral (with a cholesteric pitch 5 times larger than \textit{fd}) and 3.5 times stiffer with a persistence length of $P=9.9~\mu m$ \cite{Barry09}. For these much more rigid particles, we did not observe \textit{any} jump events. It shows the importance of chirality (setting the density of intrinsic defects in the hexatic columnar mesophase) and of stiffness to the dynamics of the particles. This behavior is supported by our simulations, where decreasing jumping frequency with increasing stiffness is observed (see Supplementary Material \cite{SM}).

The distribution of jump amplitudes presented in Fig. \ref{Fig_histogram}(b) shows clear maxima around jumps of a half and full particle length. This result for the columnar phase contrast with those found earlier for the smectic phase. The jump distributions in the smectic phase were sharply peaked around one rod length corresponding to the layer spacing \cite{Lettinga07,Pouget11}, and no measurable changes in the virus orientation were observed.
Limited optical resolution does not allow to discriminate displacements on the length scale of the intercolumnar spacing (about 17~nm at 130 mg/mL \cite{GreletPRL08}). Hence, we cannot determine if jump events occur due to {\it intra}-columnar displacements or due to {\it inter}-columnar displacements.

Plausible explanations for the full jumps are particle overtaking events and for the half jumps re-equilibration of a column following a particle moving out of the column. In order to verify this, we performed Brownian dynamics simulations of two toy models. The rod-like virus particles are represented by strings of beads interacting via rigid harmonic bond and bending potentials. The strength of the bending potentials are chosen to emulate very rigid and slightly less rigid particles with a contour length over persistence length of $L/P=0.09$ and $0.31$, which correspond to the $L/P$ values of {\it fdY21M} and \textit{fd} viruses, respectively. The aspect ratio $L/\sigma=5.8$ of our simulated particles is much smaller than that of the bare viruses ($133$) but motivated by computational expedience/restrictions. Non-bonded beads interact via a purely repulsive shifted Lennard-Jones potential.
All the Brownian dynamics simulations are performed using the LAMMPS code \cite{lammps}. In our simulations, the self-diffusivity of a single bead equals $D_b = \sigma^2/t^{\ast}$, with $\sigma$ and $t^{\ast}$ the diameter of a single bead (the unit of length) and the unit of time in the simulations. In the free-draining limit of our simulations, the self-diffusivity of our elongated particles of $n$ beads reads $D=D_b / n$ (we specify $n$ below) \cite{Doi86}.

As advertised, one explanation for the full jumps may be that of particles overtaking within a column. To test this, we first simulate a single column of particles subject to a cylindrically symmetric Gaussian confinement potential, mimicking the self-consistent molecular field that the particles experience in the columnar phase. The strength of this confining potential that is applied to all beads, $k$, is chosen to match the calculated strength of the molecular field from the simulations of Belli and collaborators referred to earlier \cite{Belli10}. Initially, the particles are placed equidistantly on the center line of the confining potential. Their linear fraction is $\psi = NL / \Lambda$, with $N$ the number of particles, $L$ their length and $\Lambda$ the length of the simulation box. We have chosen a high linear fraction of $\psi=0.90$ expected for the dense columnar phase.

\begin{figure}
\includegraphics[width=0.45\textwidth]{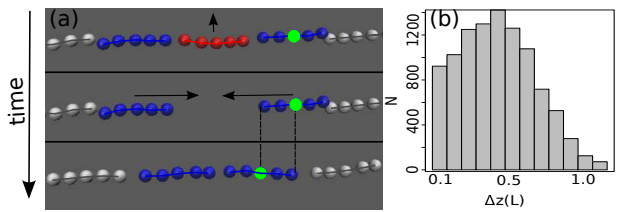}
\caption{\label{Fig_half_jump_sim} (a) Three snapshots of a simulation, where a particle (colored in red) is removed from the middle of a quasi-one-dimensional system and the remaining particles re-equilibrate. Displacement of the two neighbors (colored in blue) of the removed particle is measured after the particle is removed. (b) Probability distribution of displacement $\Delta z$ of the two neighbors after re-equilibration. Here, each elongated particle consists of five beads with an equivalent aspect ratio of $5.8$.}
\end{figure}

In order to study the influence of boundary conditions and system size, we carry out simulations for 10 and 200 particles and impose reflective and periodic boundary conditions along the main axis. With the latter we emulate an infinitely large system and with the former the poly-domain structure of the columnar phase of \textit{fd} virus, where the size of each domain is estimated to be of the order of 10 rod lengths in the axial direction. We extracted the traces of the centers-of-mass of the particles from the simulations. For periodic boundary conditions the traces are very noisy and do not display sudden changes in the positions (data not shown, see Supplementary Material \cite{SM}). For reflecting boundary conditions, we do find that individual particles suddenly change position in a manner similar to what we observed in our experiments, except that they are full jumps. The effect of the boundary conditions must be due to the fact that reflecting boundaries cause positional ordering of the particles near the boundaries, suppressing fluctuations in the position of the particles. We come back to this below.

\begin{figure}
\includegraphics[width=0.45\textwidth]{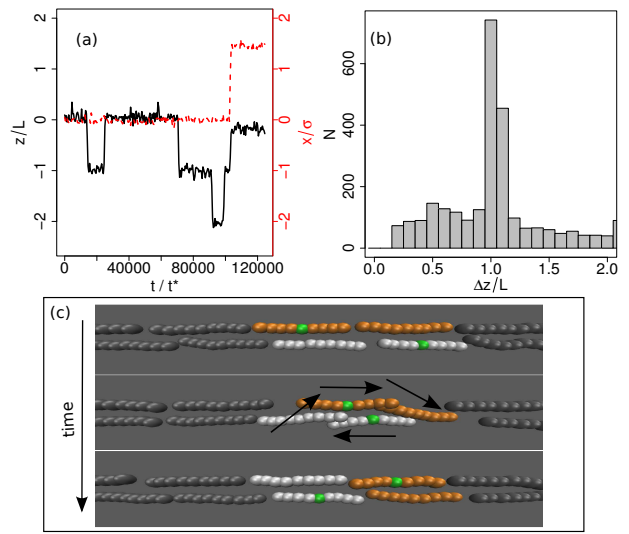}
\caption{\label{Fig_trace_sim} (a) Displacement of a rod particle of length $L$ and diameter $\sigma$ along ($z$, black solid line) and perpendicular to ($x$, red dashed line) the main axis of a column. Simulations with reflecting boundary conditions and 100 columns with an overall packing fraction of 28\%. In each column there are $N=12$ particles corresponding to a linear fraction of $\psi=0.95$.
%The strength of the confining potential equals $k = 1.5 k_BT / \sigma^2$.
 Each elongated particle consists of nine beads with an equivalent $L/P=0.31$. %and an aspect ratio of $5.8$.
 (b) Distribution of the jump length, $\Delta z$. (c) Snapshots of two columns at three different times. Particles are colored orange and white according to their initial position. The images show a circular-type movement with inter- and intra- (particles marked by a green bead) columnar jumps. Gray particles are not involved in this movement.}
\end{figure}

The full jumps in the quasi one-dimensional simulations are due to particles overtaking each other. To model half jumps we remove a particle from the middle of a column. The strength of the confining potential is chosen such that particles do not overtake each other in the simulation time window. We then measure the displacement of its neighboring particles as a function of time (Fig. \ref{Fig_half_jump_sim}(a)). To obtain sufficient statistics we repeated these simulations 10000 times. This results in the histogram of jump lengths shown in Fig. \ref{Fig_half_jump_sim}(b) for the case of $N=11$ particles and linear fraction $\psi=0.95$. The strength of confinement also allows for a direct comparison of the speed at which the jump takes place with the Tonks model that we present in the Supplementary Material \cite{SM}.

As expected, the histogram peaks at $x=L/2$ (Fig. \ref{Fig_half_jump_sim}(b)). If we extend this analysis for pairs of next-, next-next- and so on nearest neighbor, then the peak in the histogram increasingly shifts towards smaller values than one-half. If we average over all pairs, the peak in the histogram is located at zero distance. Hence, our simulations imply that the particles involved in the half jumps must be the neighbors of the particle that was removed from the column and hence that the domain size in the direction of the long axis is small. Given the fact that one fifth of the rods undergo a strong angular displacement, the estimate of a domain-size of about 5 rod \textit{lengths} seems reasonable.

We now have two scenarios that provide a rationalization of full and half jumps albeit that they are sensible only in sufficiently small systems. To study other scenarios that involve motion of particles \textit{between} columns, we invoke our second toy model in which we \textit{construct} a three-dimensional columnar phase. This is necessary because columnar phases of elongated particles are extremely difficult to obtain in a simulation without invoking length polydispersity or an external alignment field in combination with a soft repulsive interaction potential \cite{wensink07,Cinacchi09}.

To generate a columnar phase, we use two ingredients to suppress the smectic phase that turns out to be the equilibrium phase for our rod particles. First, a superposition of three periodic potentials is switched on to keep the particles to the centers of the columns producing a triangular lattice. The periodicity of the potentials fixes the lattice spacing $d$ and their amplitude $k$ the level of confinement of particles within the columns, and through that the exchange of particles between them. Second, to prevent correlated layering to occur near the reflecting boundaries of the columns, we randomly displace each column by up to a full particle length drawn from a uniform distribution. Our simulation box consists of 100 columns and we apply periodic boundary conditions in the direction perpendicular to them. We again choose a high linear fraction of particles $\psi=0.95$ in each column. The total number of rods varied from 1200 to 20000 depending on system size.

From our 3D simulations, we find another interesting scenario for the full jumps, involving fast correlated motion of collections of particles in an otherwise more or less static configuration of the majority of particles. We rarely observe intra-columnar overtaking of rods. This is in itself not surprising given the small inter-columnar distance of $1.75\sigma$, which is similar to the experimental value. Instead, we find that multiple inter-columnar jumps occur, where rod particles jump into neighboring columns at different positions (Fig. \ref{Fig_trace_sim}(a)). This leads to a collective movement of particles in two or three or four and so on columns, involving on average four particles per column. The particles involved in this movement engage mainly in a full jump (Fig. \ref{Fig_trace_sim}). %The jump length distribution has a peak at one particle length as shown in Fig. \ref{Fig_trace_sim}(b).
It should be noted (i) that this kind of event is extremely rare and (ii) that a similar kind of correlated motion is observed in disordered, glassy colloidal systems \cite{Kegel2000}. The reason of the very low frequency of the collective string-like motion is that inter-column movement requires overcoming a free-energy barrier. This barrier must be lower for flexible particles than for rigid ones because they can adjust their shape and minimize exposure to molecular field. This means that at the same density, more flexible rods should exhibit more full and half jumps than rigid ones. This is exactly what we see both in our experiments and in our simulations for two different rod flexibilities, corresponding to $L/P=0.09$ (\textit{fdY21M}) and $0.31$ (\textit{fd}). In the simulations, the jump frequency goes down by a factor of three for the more rigid particles (See Supplementary Material \cite{SM}).
Clearly, our toy model describes only part of what is observed in our experiments. For instance, in our simulations the jumping particles cannot reorient whereas in our experiments about a \textit{fifth} do (Fig. \ref{Fig_histogram}(a)). This is presumably associated with particles jumping from one domain to another one with different orientation.

In conclusion, filamentous virus particles undergo full- and half-jump displacements in the columnar phase in aqueous solution. Our simulations show that the finite domain structure plays a crucial role since half jumps are only observed for sufficiently small domains after creation of a void. The presence of such voids are corroborated by the fact that a fraction of the virus particles undergo a significant angular displacement. Full jumps could be either due to overtaking particles  within columns or to rearrangements within neighboring columns resulting from particles jumping between them. The former seems unlikely in practice unless assisted by breathing modes of the hexagonal lattice, in which case one might speak of phonon-assisted hopping.
More general, our combined experimental and simulation approach shows that heterogeneous dynamics is not limited to glassy systems but may exist in highly ordered systems too. Moreover, finite-size effects are not only relevant in computer simulations but in actual experimental systems as well.

\begin{acknowledgments}
This project was supported by the European network of excellence SoftComp. The work of S.N. forms part of the research program of the Dutch Polymer Institute (DPI, Project No. 698).
\end{acknowledgments}

\end{document}